\documentclass[aps,twocolumn]{revtex4-1}
\usepackage{amsthm, amsmath, amssymb}
\usepackage{enumerate} 
\begin{document}
 
\title{Exact solutions in metric $f(R)$-gravity for static axisymmetric spacetime}
 
\author{Antonio C. Guti\'errez-Pi\~{n}eres}\email[Email address:]
{acgutierrez@correo.nucleares.unam.mx}

\affiliation{Facultad de Ciencias B\'asicas, Universidad Tecnol\'ogica
 de Bol\'ivar, CO 131001 Cartagena de Indias, Colombia}
\affiliation{Instituto de Ciencias Nucleares,
Universidad Nacional 
Aut\'onoma de M\'exico,
A.P. 70-543, 04510 M\'exico D.F., M\'exico}

\begin{abstract}
Axially symmetric static vacuum exact solution (ASSVES) in Weyl coordinates are studied in
$f(R)$ theories of gravity. In particular, we obtain a general explicit expression for the
dependence between $df(R)/dR$ and the $r$ and $z$ coordinates and then the corresponding
explicit form of $f(R)$ is obtained. Next, we present in detail the explicit solution of
the modified field equations corresponding to the Newtonian potential due to a punctual
mass placed at the origin of coordinates and also the Schwarzschild solution to the
modified field equations. 
\end{abstract}

\maketitle

\section{Introduction}
\label{sec:intro}

In the last years, $f(R)$-gravities as modified theories have much attentions
as one of promising  candidates for to overcome the so-called cosmological
constant  or dark  energy problem (see \cite{RevModPhys.82.451} for recent
reviews). Thus, there has been a wide physical and  mathematical stimulus
for their study, leading to a strong number of interesting results in the 
context  of the exact solutions.  

However, to find exact solutions in $f(R)$-gravities is a very difficult problem
due to the highly  nonlinearity of the equations involved. One have to impose
some symmetry in order to simplify the problem. Spherically
symmetric solution are the most widely studied exact solutions in the context of
$f(R)$-gravity \cite{PhysRevD.74.064022} mainly due to the solvability of
the equations and also more importantly due to the astrophysical interest
present in this kind of symmetry. Also in the recent years some exact
solutions for static cylindrically symmetric in $f(R)$-gravity are presented
\cite{Azadi2008210}. Now, although there has been a lot of  work in the last
years we think it  would be fair to say that, except for maybe
\cite{0264-9381-27-16-165008}, we  do not have  a fully integrated explicit
exact axially symmetric solution of $f(R)$-gravity.

Our purpose here  is to consider axially symmetric static vacuum solution in
Weyl coordinates  in $f(R)$ theories of gravity. In particular, we obtain the
general explicit expression for the dependence between $df(R)/dR$ and the $r$
and $z$ coordinates and then the corresponding explicit form of $f(R)$ is
obtained. Next, we present in detail the explicit solution of the modified field
equations corresponding to the Newtonian potential due to a punctual mass placed
at the origin of coordinates and also the Schwarzschild solution to the modified
field equations.

The  paper is organized  as  follows. In section \ref{sec:FE}, the action of $f(R)$
gravity is introduced and the respective  gravitational field equations for static axially
symmetric space-time are presented. Next, in section \ref{sec:ES},  we  restrict the
problem by  considering the vacuum  only, and, using the  widely known  method of
separation of variables,  the general explicit expression for the dependence between
$df(R)/dR$ and the $r$ and $z$ coordinates and the corresponding explicit form of $f(R)$
are obtained. Then, in section \ref{sec:part} we present in detail the explicit solution
of the modified field equations corresponding to the Newtonian potential due to a punctual
mass placed at the origin of coordinates and also the Schwarzschild solution to the
modified field equations.% Conclusion are drawn in section \ref{sec:conclude}.

\section{\label{sec:FE} Field equations in $f(R)$ gravity for static  axially symmetric
space-time }
The action for $f(R)$ gravity is given by 
\begin{eqnarray}
 S=\int{\left(\frac{1}{16\pi G}f(R) + {\cal L}_{\text
m}\right)\sqrt{-g}d^4x},\label{eq:action}
\end{eqnarray}
where $G$ is the gravitational  constant, $R$ is  the  Ricci (curvature) scalar and
${\cal L}_{\text
m}$ is  the matter Lagrangian.
The   field equation  resulting from  this  action are 
 \begin{equation}
 G_{\mu\nu} = R_{\mu\nu} - \frac{1}{2}Rg_{\mu\nu} = 8\pi G(\tilde
T_{\mu\nu}^{\text{g}}
+ \tilde T_{\mu\nu}^{\text m}),\label{eq:fe}
 \end{equation}
where  the gravitational stress-energy tensor is 
\begin{eqnarray}
 8 \pi G \tilde T_{\mu\nu}^{\text g} = T_{\mu\nu}^{\text g}
= \frac{1}{F_R}\Big(
\frac{1}{2}g_{\mu\nu}\Big(f(R) - RF_R\Big) 
\\+ F_R^{;\alpha\beta}(g_{\alpha\mu} g_{\beta\nu} - g_{\mu\nu} g_{\alpha\beta})
\Big),\nonumber
\end{eqnarray}
with $F_R\equiv df(R)/dR$ and $\tilde T_{\mu\nu}^{\text
m}\equiv T_{\mu\nu}^{\text m}/F_R$, being $T_{\mu\nu}^{\text m}$ the energy-stress tensor
derived from the matter Lagrangian ${\cal L}_{\text m}$ in the  action (\ref{eq:action}).
We can write (\ref{eq:fe}), equivalently, in the  form
\begin{equation}
 F_R R_{\mu\nu} - \frac{1}{2}f(R)g_{\mu\nu} - \nabla_{\mu}\nabla_{\nu}F_R
+ g_{\mu\nu} \Box F_R = 8\pi G  T_{\mu\nu}^{\text m}. \label{eq:MFE}
\end{equation}
The  contraction of above field equations gives the following relation between $f(R)$  and
its derivative $F_R$:
\begin{equation}
 F_R R - 2f(R) + 3\Box F_R = 8\pi G T^{\text m}, \label{eq:contrac}
\end{equation}
where, $T^{\text m} = T_{\mu}^{\text m\mu}$ is the trace of the energy-stress.

Interested in the  static axially symmetric solutions of (\ref{eq:MFE}), we start  with
the Weyl-Lewis-Papapetrou metric in the cylindrical coordinates 
$x^{\alpha} = (t,\varphi,r,z)$ given by \cite{KSMH}
\begin{equation}
 ds^2= -e^{2\phi}dt^2 + e^{-2\phi}[r^2d\varphi^2 + e^{2\lambda}(dr^2
+ dz^2)]\label{eq:metric},
\end{equation}
where $\phi$ and $\lambda$ are continuous functions of $r$ and $z$.
Using (\ref{eq:contrac}), the modified Einstein equations (\ref{eq:MFE})
becomes
\begin{equation}
 F_R R_{\mu\nu} - \nabla_{\mu}\nabla_{\nu}F_R - 8\pi GT_{\mu\nu}^{\text
m}
= {g_{\mu\nu}}B, \label{eq:modified}
\end{equation}
where $B= \frac{1}{4}(F_RR - \Box F_R -8\pi G T^{\text m})$. For  the metric
(\ref{eq:metric}), the non-zero components of  the Ricci curvature tensor are simply given
by
\begin{subequations}\begin{eqnarray}
&& R_{00} = e^{4\phi -2\lambda}\nabla^2\phi, \\
&&R_{11} = r^2e^{-2\lambda}\nabla^2\phi,\\
&&R_{22} = -\nabla^2\lambda + \nabla^2\phi + \frac{2}{r}\lambda_{,r}
- 2\phi_{,r}^2,\\
&&R_{33} = -\nabla^2\lambda + \nabla^2\phi 
- 2\phi_{,z}^2,\\
&&R_{23} = \frac{1}{r}\lambda_{,z}- 2\phi_{,r}\phi_{,z},
\end{eqnarray}\label{eq:ricci}\end{subequations}
while  a straightforward computation  of the curvature
scalar of this metric leads to following result:
\begin{equation}
 R=2e^{2\phi - 2\lambda}(-\nabla^2\lambda + \nabla^2\phi +
\frac{1}{r}\lambda_{,r} - \phi_r^2 - \phi_z^2),
\end{equation}
where $\nabla^2$ is the usual  Laplace operator in cylindrical coordinates. Thus, from
(\ref{eq:modified}) and (\ref{eq:ricci}) we have the following equations system:
\begin{subequations}
\begin{eqnarray}
&&\frac{F_RR_{00} - 8\pi GT_{00}^{\text
m}}{g_{00} } = B,\\
&&\frac{F_RR_{11}- 8\pi GT_{11}^{\text
m}}{g_{11}} = B,\\
&&\frac{F_RR_{22} - 8\pi GT_{22}^{\text
m} - F_{R,22}}{g_{22}} = B,\\
&&\frac{F_RR_{33} - 8\pi GT_{33}^{\text
m} - F_{R,33}}{g_{33}} = B,\\
&&F_RR_{23} - 8\pi GT_{23}^{\text
m} -  F_{R,23} = 0.
\end{eqnarray}
\end{subequations}
The last equations system allow us to write  the following independent field equations:
\begin{subequations}
\begin{eqnarray}
\nabla^2\phi &=& -\frac{4\pi G}{F_R}e^{2(\lambda - \phi)}[T_{\;0}^{\text
m 0} - T_{\;1}^{\text m 1}],\label{eq:MEEn}\\
\lambda_{,r}&=&r(\phi_{,r}^2 - \phi_{,z}^2) \nonumber\\&+&
\frac{4\pi Gr}{F_R}e^{2(\lambda - \phi)}[T_{\;2}^{\text
m 2} - T_{\;3}^{\text m 3}] \nonumber\\
&+& \frac{r}{2F_R}[F_{R,22} -F_{R,33}],\label{eq:MEE}\\
\lambda_{,z}  &=&2r\phi_{,r}\phi_{,z}+ \frac{8\pi Gr}{F_R} T_{\;23}^{\text m}
+ \frac{rF_{R,23}}{F_R}\label{eq:MEE}.
\end{eqnarray}\label{eq:MEE}
\end{subequations}
Obviously it is  not an easy task  to find a general solution to the  above  equations, so
in the following sections we will discuss some particular solutions.

%%%%%%%%%%%%%%%%%%%%%%%%%%%%%%%%%%%%
%%%%%%%%%%%%%%%%%%%%%%%%%%%%%%%%%%%%
\section{\label{sec:ES} Vacuum solutions in $f(R)$ gravity for static  axially symmetric
space-time}
%%%%%%%%%%%%%%%%%%%%%%%%%%%%%%%%%%%%
%%%%%%%%%%%%%%%%%%%%%%%%%%%%%%%%%%%%
To simplify the problem we  suppose the vacuum case. Correspondingly, we put:
\begin{subequations}
\begin{eqnarray}
&&\nabla^2\phi =0,\label{eq:laplace}\\
&&\lambda_{,r}=r(\phi_{,r}^2 - \phi_{,z}^2) 
+ \frac{r}{2F_R}(F_{R,rr} -F_{R,zz}),\label{eq:laplacelambda1}\\
&&\lambda_{,z}  = 2r\phi_{,r}\phi_{,z}
+ \frac{rF_{R,rz}}{F_R},\label{eq:laplacelambda2}\\
&&rF_{R,z}(F_{R,rr} - F_{R,zz})
+ rF_R\nabla^2(F_{R,z})\nonumber \\
&&\qquad\qquad\;\;\; +\;\; F_{R,zr}(F_R - 2rF_{R,r}) = 0,\label{eq:integrab}\\
&&R=-2e^{2\phi - 2\lambda}(\lambda_{,rr}  + \lambda_{,zz} + 
\phi_r^2 + \phi_z^2),\label{eq:R}\\
&&f(R)=\frac{1}{2}F_RR + \frac{3}{2} \Box F_R\label{eq:f(R)},
\end{eqnarray}\label{eq:MEEvac}
\end{subequations}
being (\ref{eq:integrab}) the  condition of integrability of $\lambda$.
Otherwise, we need obtain a explicit for of $f(R)$.  Its very easy to prove that
for  the  metric
(\ref{eq:metric}) 
\begin{equation}
\Box F_R = e^{2(\phi - \lambda)}(F_{R,rr} + F_{R,zz}).\label{eq:boxF_R}
\end{equation}
So, by inserting (\ref{eq:boxF_R}) in (\ref{eq:f(R)}) and using 
(\ref{eq:R}),
we have
\begin{eqnarray}
 f(R)=\frac{1}{2}F_RR\left[1 - \frac{3W(r)}{2F_R}\right],
\label{eq:f(R)exp}
\end{eqnarray}
where
\begin{eqnarray}
 W(r)=\frac{F_{R,rr} + F_{R,zz}}{\lambda_{,rr}  + \lambda_{,zz} + 
\phi_r^2 + \phi_z^2}.
\end{eqnarray}
With the aim to obtain some solutions to the (\ref{eq:MEEvac}) equation  
systems, we assume some simplifications. We first  suppose that is possible to
write 
\begin{eqnarray}
 F_R(r,z)=U(r)V(z).
\end{eqnarray}
So,  by substituting back into (\ref{eq:integrab}) we have:
\begin{eqnarray}
 \frac{2U_{,r}(U_{,r}r-U) - 2rUU_{,rr}}{rU^2} =
\frac{V_{,zzz}V-V_{,zz}V_{,z}}{V_{,z}V}.
\end{eqnarray}
By setting each side equal to $l^2$, an arbitrary constant of  separation, we obtain
\begin{subequations}
 \begin{eqnarray}
 \frac{2U_{,r}(U_{,r}r-U) - 2rUU_{,rr}}{rU^2} =l^2\label{eq:sep1}\\
\frac{V_{,zzz}V-V_{,zz}V_{,z}}{V_{,z}V}=l^2\label{eq:sep2}.
\end{eqnarray}
\end{subequations}
We can write (\ref{eq:sep1}) as
\begin{equation}
 \frac{dM(r)}{dr} + \frac{M(r)}{r}=-\frac{l^2}{2}\label{eq:mequation},
\end{equation}
where $M(r)=U^{-1}U_{,r}$. We will now obtain a  solution of
(\ref{eq:mequation}), to do it, we suppose that is possible to write
\begin{eqnarray}
 M= M_h + M_p,
\end{eqnarray}
where $M_h$ is the  general solution of  the  homogeneous differential 
equation
\begin{equation}
 \frac{dM_h(r)}{dr} + \frac{M_h(r)}{r}=0,\label{eq:homomequation}
\end{equation}
which is
\begin{eqnarray}
 M_h= \frac{n}{r},
\end{eqnarray}
being $n$ an arbitrary constant. Whereas we can see that
$M_p = -{l^2r}/{4}$ is a particular solution of (\ref{eq:mequation}).
Consequently we have the  differential equation
\begin{eqnarray}
 M(r)=U^{-1}U_{,r}= \frac{n}{r} - \frac{l^2r}{4},
\end{eqnarray}
which solution is
\begin{eqnarray}
U(r)=cr^ne^{{-l^2r^2}/{8}},
\end{eqnarray}
with $c$ an arbitrary constant.  On the  other hand, its very  easy to prove
that 
\begin{eqnarray}
 V(z)= e^{bz}\label{eq:vsol},
\end{eqnarray}
being $b$ and arbitrary constant, is solution of (\ref{eq:sep2}) if $b$ and $l$
both satisfy the condition
\begin{eqnarray}
 bl^2=0.
\end{eqnarray}
So,we have  the following possible solutions for $F_R$:
\begin{enumerate}[(\bf i)]
 \item $b=0$ and $l=0$.
In this case 
\begin{eqnarray}
 F_R=cr^n.\label{eq:FR1}
\end{eqnarray}
 \item $b\neq 0$ and $l=0$.
In this case 
\begin{eqnarray}
  F_R=cr^ne^{bz}.\label{eq:FR2}
\end{eqnarray}
 \item $b=0$ and $l\neq 0$.
In this case 
\begin{eqnarray}
 F_R=cr^ne^{-l^2r^2/8}.\label{eq:FR3}
\end{eqnarray}
\end{enumerate}
Consequently, by substituting (\ref{eq:FR1}) (or (\ref{eq:FR2})) in
(\ref{eq:f(R)exp}) we receive 
\begin{eqnarray}
  f(R)= 2R\frac{df}{dR},
 \end{eqnarray}
which  solution can be  written as
\begin{equation}
 f(R)=kR^{1/2},
\end{equation}
being $k$ an arbitrary constant.

On the other hand, by inserting (\ref{eq:FR2}) in
(\ref{eq:laplacelambda1}) and (\ref{eq:laplacelambda2}), we  have
\begin{subequations}\begin{eqnarray}
&&\lambda_{,r}=r(\phi_{,r}^2 - \phi_{,z}^2) 
+ \frac{1}{2r}[n(n-1) -b^2r^2],\label{eq:lambdarpart}\\
&&\lambda_{,z}  = 2r\phi_{,r}\phi_{,z}
+ bn,\label{eq:lambdazpart}
\end{eqnarray}\end{subequations}
respectively. Whereas, by taking 
$F_R=cr^ne^{-l^2r^2/8}$ we obtain from  (\ref{eq:f(R)exp})
\begin{eqnarray}
 f(R)=\frac{1}{2}F_RR\left[1 - 3L(r)\right],\label{eq:f(R)3}
\end{eqnarray}
being
\begin{eqnarray}
 L(r)=\frac{(l^2r^2 - 4n)^2 - 4(l^2r^2 + 4n)}
{(3l^2r^2 + 4n -4)(l^2r^2 - 4n)}.
\end{eqnarray}
And, by inserting (\ref{eq:FR3}) in
(\ref{eq:laplacelambda1}) and (\ref{eq:laplacelambda2}), we  have
\begin{subequations}\begin{eqnarray}
\lambda_{,r}&=&r(\phi_{,r}^2 - \phi_{,z}^2) 
\nonumber\\&+& \frac{1}{32r}
[(l^2r^2-4n)^2 -4(l^2r^2 + 4n)],\label{eq:lambdarpart}\\
\lambda_{,z}  &=& 2r\phi_{,r}\phi_{,z},\label{eq:lambdazpart}
\end{eqnarray}\end{subequations}
respectively.

%%%%%%%%%%%%%%%%%%%%%%%%%%%%%%%%%%%%
%%%%%%%%%%%%%%%%%%%%%%%%%%%%%%%%%%%%
\hyphenation{va-cuum}
\section{\label{sec:part} Particular solutions  in $f(R)$ gravity for the
vacuum static axially symmetric space-time}

%%%%%%%%%%%%%%%%%%%%%%%%%%%%%%%%%%%%
%%%%%%%%%%%%%%%%%%%%%%%%%%%%%%%%%%%%
In this  section we  present two applications of the result obtained in the last 
section. We first suppose a given  metric potential and then we obtain the another metric
potential in ``presence'' of $f(R)$, by using 
$F_R=cr^ne^{bz}$,
and
$F_R=cr^ne^{-l^2r^2/8}$,
the previous results.

\subsection{The punctual mass $m$ at the origin}
The  metric  potential corresponding to a punctual mass $m$ at the origin of a coordinate
system is given by
\begin{eqnarray}
 \phi = -\frac{m}{\sqrt{r^2 + z^2}},
\end{eqnarray}
so, by using the result $F_R=cr^ne^{bz}$, we obtain for $\lambda$:
\begin{eqnarray}
 \lambda_{,r}&=& r(\phi_{,r}^2 - \phi_{,z}^2) +\frac{1}{2r}[n(n-1)-b^2r^2],\\
 \lambda_{,z}&=&2r\phi_{,r}\phi_{,z} +  bn,
\end{eqnarray}
consequently, $\lambda$ is given by
\begin{eqnarray}
\lambda&=&\frac{\tilde\lambda}{4(r^2 + z^2)^2},\\
\tilde\lambda&=& - 2m^2r^2 + (r^2 + z^2)^2\times\nonumber\\
&&\times [2n(n-1)\ln{(r)}  - b(br^2 -4nz)]\nonumber.
\end{eqnarray}

Whereas, if $F_R=cr^ne^{-l^2r^2/8}$, we have for the derivatives of $\lambda$:
\begin{subequations}
 \begin{eqnarray}
 \lambda_{,r}&=& r(\phi_{,r}^2 - \phi_{,z}^2) \\
 &+&\frac{1}{32r}[16n(n-1)  - 4(2n+1)l^2r^2 + r^4l^4]\nonumber\\
 \lambda_{,z}&=&2r\phi_{,r}\phi_{,z},
\end{eqnarray}
\end{subequations}
which solution is
\begin{eqnarray}
\lambda=\frac{\tilde\lambda}{128(r^2 + z^2)^2},
\end{eqnarray}
being
\begin{eqnarray*}
\tilde\lambda&=&[(l^2r^3 + l^2rz^2)^2 - 64m^2]r^2\\ &+& 
[64n(n-1)\ln{(r)}  - 8(2n+1)l^2r^2](r^2 + z^2)^2.
\end{eqnarray*}
\subsection{The Schwarzschild solution}
In prolate spheroidal coordinates $(x,y)$ related to the  cylindrical coordinates $(r,z)$
through the  relations
\begin{eqnarray}
r^2=m(x^2 - 1)(1 -y^2),\qquad z=mxy.
\end{eqnarray}
The Einstein vacuum equations in $f(R)$ gravity, for static axially symmetric space-time
can be cast:

\begin{subequations}
\begin{eqnarray}
\nabla^2\phi&=&0\\
\lambda_{,x}&=& \tilde\lambda_{,x} + \beta r_{,x} + \Omega z_{,x},\\
\lambda_{,y} &=&\tilde\lambda_{,y} + \beta r_{,y} + \Omega z_{,y},
\end{eqnarray}
\end{subequations}
where
\begin{eqnarray}
\phi=\frac{1}{2}\ln{\left[\frac{x-1}{x+1}\right]},\qquad
\tilde\lambda=\frac{1}{2}\ln{\left[\frac{x^2-1}{x^2-y^2}\right]}
\end{eqnarray}
are the metric potentials corresponding to the usual Schwarzschild solution in the
usual Einstein's gravity,
and
\begin{eqnarray}
\beta=\frac{r}{2F_R}(F_{R,rr} - F_{R,zz}), \qquad \Omega =\frac{rF_{R,rz}}{F_R}.
\end{eqnarray}
so, in the same way as the last case, we will obtain a expltheiricit form of
$\lambda$ by take
the different  values of $F_R$. So, in the case $F_R=cr^ne^{bz}$, we have 
\begin{subequations}
\begin{eqnarray}
\lambda_{,x}&=& \frac{x(1-y^2)}{(x^2-1)(x^2-y^2)}
+ \frac{n(n-1)x}{2(x^2 -1)} \\
&-& \frac{b^2m^2}{2}x(1-y^2) + bmny\nonumber\\
\lambda_{,y} &=&\frac{y}{x^2-y^2} - \frac{n(n-1)y}{2(1-y^2)}\\
&+&\frac{b^2m^2y}{2}(x^2 -1) +bmnx,\nonumber
\end{eqnarray}
\end{subequations}
which solution is
\begin{eqnarray}
\lambda&=&\tilde\lambda +\frac{n(n-1)}{4}\ln{[(x^2-1)(1-y^2)]}\\
&+& \frac{b^2m^2Q}{4} + bmnxy,\nonumber\\
Q&=&x^2y^2-x^2-y^2.
\end{eqnarray}
whereas in the case
$F_R=cr^ne^{-l^2r^2/8}$,
we have  for  the derivatives of the $\lambda$:
\begin{subequations}
\begin{eqnarray}
\lambda_{,x}&=& \tilde\lambda_{,x} + \frac{x}{32(x^2-1)}P, \\
\lambda_{,y}&=& \tilde\lambda_{,y} - \frac{y}{32(1-y^2)}P,\\
P&=& [16n(n-1) +l^4r^4
-4l^2(2n+1)r^2]\nonumber,
\end{eqnarray}
\end{subequations}
which solution is
\begin{eqnarray} 
\lambda&=&\tilde\lambda + \frac{n(n-1)}{4}\ln{[(x^2-1)(1-y^2)]}\nonumber\\
&+&\frac{l^2m^2}{128}[l^2m^2(Q+2) + 8(2n +1)]Q.\nonumber\\
\end{eqnarray}



%% References


%\bibliographystyle{elsarticle-num}



%\section{\label{sec:conclude}Conclude remarks}
The issue of  static and axially symmetric solutions in $f(R)$-gravity is an
important theme  in the context  of the exact solutions. In this paper, we have
presented axially symmetric static vacuum solution in Weyl coordinates in $f(R)$
theories of gravity. In particular, by the introduction of the integrability
condition of one of the metric potentials of the Weyl-Lewis-Papapetrou line
element and using the method of separation of variables we have obtained a
general explicit expression for the dependence between $df(R)/dR$ and the $r$
and $z$ coordinates and then the corresponding general explicit form of $f(R)$.
We have also presented  in detail the explicit solution of the modified field
equations corresponding to the Newtonian potential due to a punctual mass placed
at the origin of coordinates and also the Schwarzschild solution to the modified
field equations.\\

This  work is dedicated with great pleasure to Biky (M.V.R.H.) on the occasion of her 23rd
birthday.

\section*{Acknowledge} 
This work was supported in part by TWAS-CONACYT Postgraduate
Fellowship Programme.

\end{document}